# Hydration of $NH_4^+$ in Water: Bifurcated Hydrogen Bonding Structures and Fast Rotational Dynamics

Jianqing Guo,[1,2] Liying Zhou,[1,2] Andrea Zen,[3,4] Angelos Michaelides,[3,5] Xifan Wu,[6]
Enge Wang,[1,2,7,8,9,*] Limei Xu,[1,2,7,†] and Ji Chen[2,7,5,‡]

[1]International Center for Quantum Materials, Peking University, Beijing 100871, People's Republic of China
[2]School of Physics, Peking University, Beijing 100871, People's Republic of China
[3]Department of Physics and Astronomy, Thomas Young Centre and London Centre for Nanotechnology University College London, Gower Street, London WC1E 6BT, United Kingdom
[4]Department of Earth Sciences, University College London, Gower Street, London WC1E 6BT, United Kingdom
[5]Max Planck Institute for Solid State Research, Stuttgart 70569, Germany
[6]Department of Physics, Temple University, Philadelphia, Pennsylvania 19122, USA
[7]Collaborative Innovation Center of Quantum Matter, Beijing 100871, People's Republic of China
[8]Songshan Lake Materials Lab, Institute of Physics, Chinese Academy of Sciences, Guangdong 523808, People's Republic of China
[9]School of Physics, Liaoning University, Shenyang 110136, People's Republic of China

Understanding the hydration and diffusion of ions in water at the molecular level is a topic of widespread importance. The ammonium ion ($NH_4^+$) is an exemplar system that has received attention for decades because of its complex hydration structure and relevance in industry. Here we report a study of the hydration and the rotational diffusion of $NH_4^+$ in water using *ab initio* molecular dynamics simulations and quantum Monte Carlo calculations. We find that the hydration structure of $NH_4^+$ features bifurcated hydrogen bonds, which leads to a rotational mechanism involving the simultaneous switching of a pair of bifurcated hydrogen bonds. The proposed hydration structure and rotational mechanism are supported by existing experimental measurements, and they also help to rationalize the measured fast rotation of $NH_4^+$ in water. This study highlights how subtle changes in the electronic structure of hydrogen bonds impacts the hydration structure, which consequently affects the dynamics of ions and molecules in hydrogen bonded systems.

The hydration and diffusion of ions and molecules in water is one of the most fundamental processes in nature and in modern technology, having a direct impact on nucleation and crystallization, ion sieving, and aqueous chemical reactions, to name just a few examples [1–5]. A prototypical example of ion solvation with hydrogen bonding is the ammonium ion ($NH_4^+$) in water. This system has a complex and debated hydration structure. In simulations the coordination number of $NH_4^+$ is around five [6–8], whereas experiments indicated that the coordination number is larger [9]. In addition, there is an interesting but unresolved experimental observation that $NH_4^+$ rotates rapidly in water [10,11]. The estimated rotational diffusion constant from nuclear magnetic resonance measurements is a few times larger than theoretical estimates.

In 1999, Brugé *et al.* proposed a mechanism involving discontinuous rotational jumps associated with the exchange of water molecules [6]. Within this model, the $NH_4^+$ ion forms four long-lived hydrogen bonds with water, and exchange occurs between the fifth water molecule and the four water molecules bonded with $NH_4^+$ in the first hydration shell. In 2005, Intharathep *et al.* suggested that $NH_4^+$ undergoes free rotation due to a flexible hydration structure and a large coordination number [12]. However, neither of the suggested mechanisms are supported by accurate investigations of the water-ammonium ion interactions and of the underlying potential energy surface.

In order to correctly simulate the diffusion of ions in water it is essential to use a theoretical method that is able to reliably describe hydrogen bonds and the energetics involved [6,13,14]. In recent years, many breakthroughs have been reported in understanding the hydrogen bonding structure of water and the dynamics of ions in water using molecular dynamics simulations based on density functional theory (DFT) [13,15–19]. However, the outcome of DFT crucially depends on the choice of the exchange correlation functional [17]. Therefore, performing molecular dynamics simulations based on well validated exchange correlation functionals is desired. In order to validate DFT, high level theories should be applied to evaluate the interactions in the $NH_4^+$ solution. Recently, quantum Monte Carlo methods, especially fixed node diffusion Monte Carlo (DMC) calculations, have been used in high-pressure ice, bulk water, molecular crystals, and other extended systems to provide highly accurate benchmarks [20–24]. Once the questions over the bonding and the hydration structure are answered, one may be able to clarify the rotational mechanism.

In this study, we carried out *ab initio* molecular dynamics (AIMD) simulations of an $NH_4^+$ aqueous solution using three different state of the art exchange correlation functionals. We then performed DMC calculations to benchmark the accuracy of each functional and identify the most reliable functional(s). We find that DMC calculations and DFT, with the strongly constrained and appropriately normed (SCAN) functional, correctly reproduce the energy ordering of the hydration structures, where the most favorable coordination consists of six water molecules in the first hydration shell, which is in good agreement with experimental indications. In the stable hydration structure, multiple water molecules appear at the so-called bifurcated positions (a bifurcated position is a position between two protons of $NH_4^+$), forming bifurcated nonlinear hydrogen bonds (a nonlinear hydrogen bond has a large hydrogen bond angle). Such a bifurcated hydration structure directly leads to the fast rotation of $NH_4^+$ in water, and the rotational diffusion constant calculated is in good agreement with experimental measurements [11]. We also find that there is a relation between the rotational diffusion constant and the number of bifurcated hydrogen bonds predicted with different exchange correlation functionals.

DFT-based AIMD simulations were carried out using the Quantum ESPRESSO package [25,26] with the SCAN [27], the Perdew-Burke-Ernzerhof (PBE) [28], and the PBE + vdW [29] functionals. The interactions between the valence electrons were treated with Hamann-Schlüter-Chiang-Vanderbilt pseudopotentials [30,31]. Each simulation was performed in the canonical (NVT) ensemble, and the temperature was controlled by a Nosé-Hoover thermostat [32]. Hydrogen atoms were replaced with deuterium in order to reduce nuclear quantum effects and to maximize the time step in the integration of the equations of motion. DMC calculations were performed using the CASINO package [33], with the size-consistent DMC algorithm ZSGMA [34]. The recently developed energy consistent correlated electron pseudopotentials [35] were used. To reduce or eliminate the finite size error (FSE), we used the model periodic Coulomb (MPC) interaction method [36–38]. In previous studies, the above setup has been validated for noncovalent interactions through extensive comparisons with converged coupled cluster calculations and experimental evaluations [22,39,40]. Further computational details can be found in the Supplemental Material [41], which includes additional Ref. [42].

We begin our study by discussing the hydration structures from AIMD simulations, as obtained using SCAN, PBE, and PBE + vdW. Figure 1(a) shows the N-O radial distribution function (RDF) $G_{NO}(r)$ and the corresponding running coordination number (CN) $n_{NO}(r)$, where CN is defined as the average number of water molecules in the first hydration shell. The first peak in $G_{NO}(r)$ with SCAN is higher and broader than those with PBE and PBE + vdW, leading to a significantly larger CN of around 6.6, while the results of PBE (CN = 4.5) and PBE + vdW (CN = 5.4) are smaller. The distribution of CNs shown in Fig. 1(b) further demonstrates the differences of CN between SCAN and the two other functionals. The most favorable coordination number of SCAN is six followed by seven and five, while the CN distribution of PBE and PBE + vdW peak at four and five, respectively. As for the second hydration shell, PBE and PBE + vdW both have a clear second peak on $G_{NO}(r)$, whereas with SCAN the second hydration shell mixes with the first shell. The flattening of the second peak predicted by SCAN has been observed in neutron diffraction and x-ray experiments [9,43].

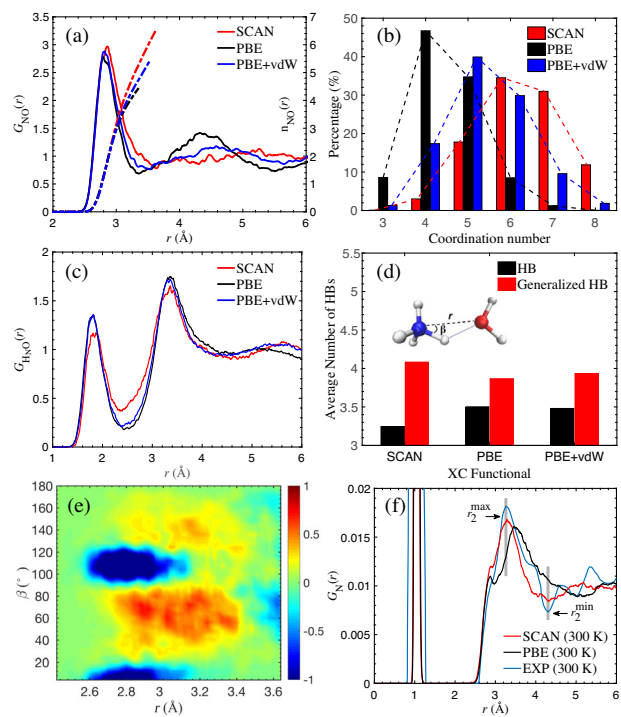

FIG. 1. Hydration structures of $NH_4^+$ in water obtained from AIMD. (a) The N-O radial distribution function (solid line) and the corresponding running coordination number (dashed line). The running coordination number $n_{NO}(R) = 4\pi\rho \int_0^R G_{NO}(r) r^2 dr$, where $\rho$ is the density. The first minimum ($r_{min}$) of $G_{NO}(r)$ is 3.64 Å, 3.38 Å, 3.55 Å for SCAN, PBE, PBE + vdW, respectively. The coordination number (CN) is $n(r_{min})$. The convergence of $G_{NO}(r)$ is shown in Fig. S2(a). (b) The distribution of the coordination number of the first hydration shell. (c) Calculated $H_N$-O radial distribution functions. The first minimum of $G_{H_NO}(r)$ is 2.40 Å, 2.45 Å, 2.39 Å for SCAN, PBE, PBE + vdW, respectively. (d) Average number of $NH_4^+$-water hydrogen bonds (HBs) predicted by different exchange correlation (XC) functionals. (e) The difference of the joint distribution in the first hydration shell between the SCAN and the PBE results as a function of $r$ and $\beta$ [defined in the inset of (d)]. Red color implies a larger value from SCAN. The joint distributions are shown in Fig. S3. (f) The N-water total distribution function $G_N(r)$. The blue solid line is from the neutron diffraction work of Hewish *et al.* [43]. The grey bars indicate the positions of the second maximum and the second minimum of the experimental $G_N(r)$.

We now consider the number of hydrogen bonds formed by $NH_4^+$ and the surrounding water molecules. Figure 1(c) plots the $N_N$-O radial distribution functions from AIMD. We define the standard hydrogen bonds with criteria reported for liquid water and aqueous solutions [44–46], i.e.,

$$R^{NO} < R_C^{NO}, \qquad R^{H_NO} < R_C^{H_NO}, \qquad \beta < 30°, \quad (1)$$

where $R_C^{NO}$ and $R_C^{H_NO}$ are the cut-off values for the N-O and $H_N$-O distances obtained from the first minimum of the corresponding radial distribution functions. The angle $\beta$ is the $H_N$-N-O angle as shown in the inset of Fig. 1(d). We find the number of $NH_4^+$-water hydrogen bonds predicted by PBE and PBE + vdW are almost identical with an average value of 3.50 and 3.48, respectively, suggesting a minor influence of vdW interactions on the $NH_4^+$-water hydrogen bonds. SCAN, however, predicts considerably weaker $NH_4^+$-water hydrogen bonds evidenced by a smaller average number of 3.24. By ignoring the angular restraint in Eq. (1), i.e., one can define the so-called generalized hydrogen bonds, including nonlinear, bifurcated hydrogen bonds that have $\beta > 30°$ [47,48]. We find that SCAN predicts a larger number of bifurcated $NH_4^+$-water hydrogen bonds.

The hydration structure of $NH_4^+$ in water can also be demonstrated by the joint distributions of $r$ and $\beta$, which are plotted in Fig. S3 [41]. The peaks at $\beta < 30°$ and $\beta \approx 100°$ represent the water molecules that form hydrogen bonds with $NH_4^+$ and the increased intensity in between can be attributed to the bifurcated water molecules that tend to form multiple nonlinear hydrogen bonds with the $NH_4^+$. To highlight the difference between different AIMD simulations, in Fig. 1(e) we plot the difference of the joint distribution of SCAN and PBE. Because of the weaker $NH_4^+$-water hydrogen bonds and increased bifurcated water molecules predicted by SCAN, the peaks at $\beta < 30°$ and $\beta \approx 100°$ are reduced while the intensity in the middle becomes stronger.

Figure 1(f) further plots the $N$-water total distribution function [$G_N(r)$] obtained from our simulations and from available neutron diffraction data. Experimental $G_N(r)$, as defined in Eq. (2), is different from $g_{NO}(r)$ in Fig. 1(a) and involves $g_{NN}$ and $g_{NCl}$ terms because experiments used 5.0 mol/L $NH_4Cl$ water solutions [6,9].

$$G_N(r) = 0.00260 g_{NO}(r) + 0.00716 g_{NH}(r) \\ + 0.00043 g_{NCl}(r) + 0.00035 g_{NN}(r) - 0.01038. \quad (2)$$

Our simulated $G_N(r)$, however, contains only the first two terms in Eq. (2). The difference in solution conditions is apparent from the difference between curves at large radii. Nevertheless, at small radii, the comparison indicates that the SCAN result is reasonable, in particular the position of the second maximum $r_2^{max}$ and the second minimum $r_2^{min}$ are in line with the experiment. Simulations with an inaccurate functional such as PBE do not correctly reproduce these experimental features. In addition, Figs. S4 and S5 [41] present vibrational spectra and density of states computed from our simulations, which may be compared in future experiments to confirm the proposed hydration structure.

Overall, we find that the hydration structure predicted by SCAN is more consistent with experiments than that obtained with the other two functionals, with the SCAN results characterized by weaker $NH_4^+$-water hydrogen bonds and an increased number of water molecules in the first hydration shell. However, to confirm that the SCAN description of the hydration structure is indeed reliable and the agreement between theory and experiment is not merely fortuitous, it is necessary to examine the accuracy of the underlying potential energy surface with the help of a higher level of theory. To this end we selected five distinct configurations with CN from four to eight to benchmark against DMC calculations, as shown in Figs. 2(a–e). Relative energies of these configurations are presented in Fig. 2(f). According to DMC, the configuration with CN = 6 is the lowest energy state among the selected structures. The basin shape of the DMC predicted energy curve is reproduced quite well by SCAN, indicating a good description of the hydration structure and the hydrogen bonding network around $NH_4^+$. Indeed, the energy ordering of the selected structures predicted by DMC and SCAN is qualitatively consistent with the distribution of CNs shown in Fig. 1(b). Regarding the performance of other DFT functionals, typical results are shown with PBE and PBE + vdW. From this it can be seen that they predict quite different energy ordering for the same set of structures. Specifically, both PBE and PBE + vdW tend to underestimate the stability of the CN = 6 configuration. Furthermore, SCAN correctly predicts that the configuration of CN = 6 is the lowest energy state while PBE and PBE + vdW both overestimate the stability of the state of CN = 5. In Fig. S6 [41], we show the results calculated using several other exchange correlation functionals and none of them predict a better energy ordering. Benchmarks on the gas phase $NH_4^+$-water clusters presented in Fig. 2(g) further support our conclusions. It is worth noting that recent experimental measurements also suggested that hydrogen bonds of solvated $NH_4^+$ are less strong than previous theoretical predictions [8]. Here our DMC and DFT calculations clarify that the $NH_4^+$-water interaction is indeed weaker than previously thought, and the relatively weak interaction leads a large coordination number of $NH_4^+$ and an increased number of nonlinear hydrogen bonds. Theoretically, the relatively weak hydrogen bonds predicted by SCAN can be understood by the lower polarizability of the water molecules around $NH_4^+$ (Fig. S5); a suggestion that is consistent with previous studies in water [18,19].

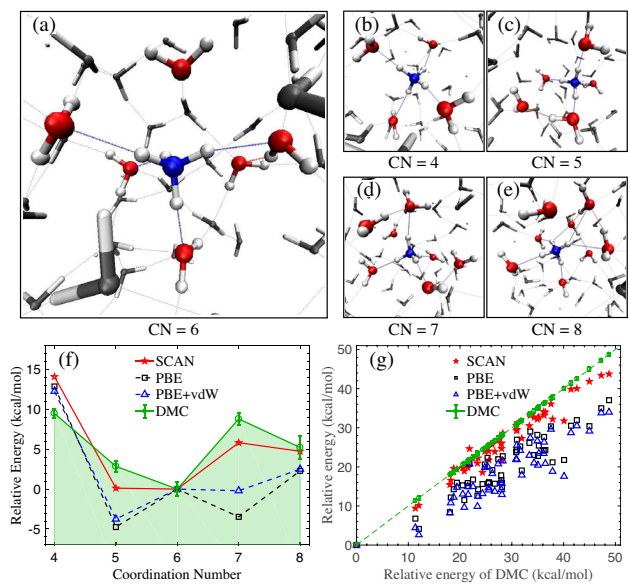

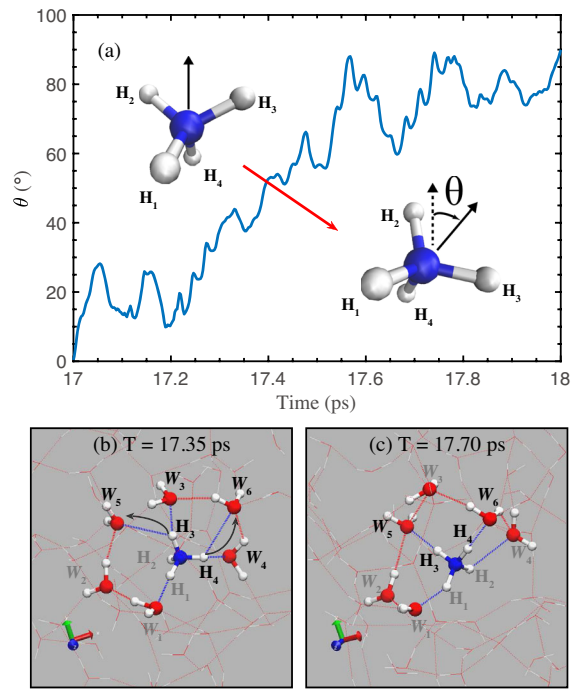

FIG. 2. Benchmarks of different DFT functionals against DMC calculations. (a)–(e) Snapshots of the selected configurations. Atoms in bright colors represent the first hydration shell of $NH_4^+$ in water (blue: N; red: O; white: H). (f) Relative energies calculated using different methods at different configurations with CN from four to eight. The relative energy is defined as the energy difference to the energy of the configuration with CN = 6. (g) Relative energies of gas phase $NH_4^+$ clusters calculated using different methods and plotted against the DMC results. The relative energy is defined as the energy difference to the energy of the configuration with the lowest one. The gas phase $NH_4^+$ clusters are randomly selected from the AIMD trajectories, which consist of $NH_4^+$ and eight neighboring water molecules.

FIG. 3. Rotation mechanism of $NH_4^+$ in water. (a) Rotation angle $\theta$ (the angle change with respect to $T = 17.0$) as a function of time. Calculated $\theta$ is averaged by the vector $\overrightarrow{NH_2} + \overrightarrow{NH_3}$, $\overrightarrow{NH_2} + \overrightarrow{NH_4}$, and $\overrightarrow{NH_3} + \overrightarrow{NH_4}$. (b)–(c) Snapshots of a typical rotational process. The blue sphere is N, red spheres are O, and white spheres are H. $H_1$ to $H_4$ belongs $NH_4^+$. Water molecules $W_1$ to $W_6$ represent the first hydration shell at $T = 17.0$ ps. Rotation is associated with bifurcated hydrogen bonds formed with $H_3$ and $H_4$.

Having established that the hydration structure of $NH_4^+$ is characterized by weak and bifurcated hydrogen bonds, and having found that the SCAN functional can reproduce this structure, we now discuss the rotation of $NH_4^+$ in water. To aid the quantification of the rotation, we define a vector in the body-fixed frame and track the rotation of the vector as a function of time. The angle $\theta$ is defined as the angle between the initial vector and the vector after a given time. Figure 3(a) shows a typical evolution of $\theta$ in a picosecond window when the rotation of $NH_4^+$ occurs, which is described by a big change of $\theta$ of 80–90° in 0.3–0.4 ps. The details of the rotation are described in Fig. S7 and Movie S1 of the Supplemental Material [41].

Figures 3(b)–3(c) show two snapshots just before and after the jump to highlight the key step of the rotation, namely the simultaneous switching of two bifurcated hydrogen bonds. In Figs. 3(b) and 3(c), $W_1$ to $W_6$ are the six water molecules in the first hydration shell of $NH_4^+$ at $T = 17.0$ ps. Initially two bifurcated water molecules ($W_5$ and $W_6$) form nonlinear hydrogen bonds with $NH_4^+$, hence $H_3$ and $H_4$ form bifurcated hydrogen bonds with $W_3/W_5$ and $W_4/W_6$, respectively. After the simultaneous switching of the two bifurcated hydrogen bonds, N-$H_3$ forms a stable hydrogen bond with $W_5$ and N-$H_4$ points towards $W_6$. Because of the bifurcated hydrogen bonds, such a mechanism does not involve complete breaking of hydrogen bonds, hence facilitating the rotation. In addition, the rotation of $NH_4^+$ involves the rotation of two N-H bonds instead of one, thus it is much more efficient when there is a pair of bifurcated hydrogen bonds. A statistical analysis of rotation times and the extent of the rotational angle jumps is provided in Fig. S8, confirming that large rotational jumps dominate the overall rotational process. In contrast, with PBE and its hydration shell containing fewer bifurcated water molecules, the rotation involves the breaking of hydrogen bonds, and the process is slowed down (Fig. S9).

The above rotational process is characterized by the rotational diffusion constant $D_R$ [7,12], which can be derived from the mean-square angular displacements according to

$$\langle \theta(t)^2 \rangle = 4D_R t. \tag{3}$$

The computed $\langle \theta(t)^2 \rangle$ as a function of time $t$ are shown in Fig. S10(a) [41]. From the linear part of $\langle \theta(t)^2 \rangle$ curve, we can estimate the rotational diffusion constant $D_R$ which is a quarter of the slope. The estimate for $D_R$ using SCAN is $0.156 \pm 0.009$ rad$^2$ ps$^{-1}$ at 363 K (the convergence of $D_R$ was shown in Fig. S11) while the values from PBE $(0.050 \pm 0.007)$ and PBE + vdW $(0.073 \pm 0.007)$ simulations are significantly smaller. In Fig. 4(a) we plot $D_R$ as a function of temperature along with the experimental values derived from the NMR measurements [7,10–12]. We find that $D_R$ follows a good linear relation in the temperature regime studied, and a linear fit is highlighted with the purple dashed line for $NH_4^+$ in $H_2O$. Through the experimental data of $ND_4^+$ in $D_2O$ we draw a green dashed line parallel to the purple line, which suggests that our simulated $D_R$ with the SCAN functional falls in line with experiments. We note that to explicitly discuss the difference between the hydrogenated and deuterated systems, nuclear quantum effects should be included using, e.g., path integral molecular dynamics in the future [49,50]. Nevertheless, from the experimental measurements it is evident that nuclear quantum effects have a minor impact, of approximately 0.02 rad$^2$ ps$^{-1}$, on the rotational diffusion constant [11].

To quantitatively demonstrate the correlation between hydration structure and rotational dynamics, we further plot $D_R$ at 363 K as a function of the average number of bifurcated hydrogen bonds in Fig. 4(b). We find $D_R$ grows as the number of bifurcated hydrogen bonds increases, following a linear relation. Similar results are obtained at other temperatures, which are presented in Fig. S10 [41]. The linear relation established offers strong evidence for our finding that the rotation involves the simultaneous switching of bifurcated hydrogen bonds.

To conclude, we have performed AIMD simulations and DMC calculations of the $NH_4^+$ aqueous solution. We have clarified the hydration structure of $NH_4^+$ in water, finding that it is characterized by a coordination number of approximately six, with two pairs of bifurcated hydrogen bonds. Consequently, we find that this bonding nature leads to fast rotation of $NH_4^+$ in water. A clear prediction from this study is that the rotation of $NH_4^+$ may be significantly suppressed in a different solvent where bifurcated hydrogen bonds do not form or when it is confined to an extent that the hydration structure is partially broken. In the future, such studies are desirable to verify our conclusions and confirm the diffusion mechanism suggested here. Last but not the least, this study brings attention to the importance of achieving accurate electronic structures of other aqueous and hydrogen bonded systems, and further improvements in computational methods will gradually bring us towards an exact description of such complex processes.

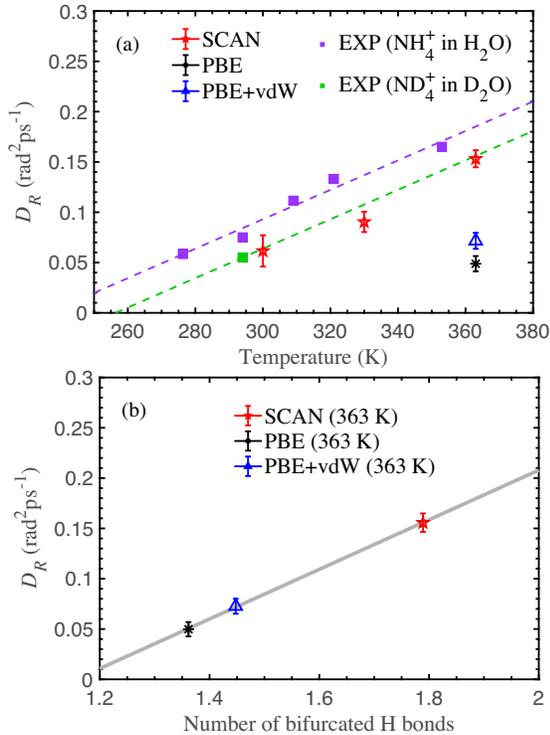

FIG. 4. Diffusion of $NH_4^+$ in water. (a) Rotational diffusion constant $D_R$ from experiments and our simulations at different temperatures [11]. The purple dashed line is a linear fit of the experimental rotational diffusion constant as a function of temperature for H, and the green dashed line is a parallel line to the purple dashed line crossing the only available experimental data for D as a guide to the eye. (b) $D_R$ calculated from our simulations at 363 K as a function of the average number of bifurcated hydrogen bonds. The gray line is a linear fit.

This work was supported by the National Key R&D Program of China under Grant No. 2016YFA030091, the National Natural Science Foundation of China under Grants No. 11974024, No. 11525520, and No. 11935002. A. Z.'s work is sponsored by the Air Force Office of Scientific Research, Air Force Material Command, U.S. Air Force, under Grant No. FA9550-19-1-7007. X. W.'s work was conducted within the Computational Chemical Center: Chemistry in Solution and at Interfaces funded by the DoE under Award No. DE-SC0019394. We are grateful for computational resources provided by TianHe-1A and TianHe II supercomputers, the High Performance Computing Platform of Peking University, and the Platform for Data Driven Computational Materials Discovery of the Songshan Lake Materials Lab.

*egwang@pku.edu.cn
†limei.xu@pku.edu.cn
‡ji.chen@pku.edu.cn